\documentclass[reprint,amsmath,amssymb,aps]{revtex4-2}

\usepackage{graphicx}

\begin{document}

\preprint{APS/123-QED}

\title{Tailoring Frictional Properties of Surfaces Using Diffusion Models}

\author{Even Marius Nordhagen}
 \email{evenmn@mn.uio.no}
\author{Henrik Andersen Sveinsson}%
\author{Anders Malthe-Sørenssen}
 \altaffiliation[Also at ]{The Njord Center}
\affiliation{%
 Department of Physics,\\ University of Oslo,\\
 Sem Sælands vei 24, NO-0316 Oslo, Norway
}%


\date{\today}

\begin{abstract}
This Letter introduces an approach for precisely designing surface friction properties using a conditional generative machine learning model, specifically a diffusion denoising probabilistic model (DDPM).
We created a dataset of synthetic surfaces with frictional properties determined by molecular dynamics simulations, which trained the DDPM to predict surface structures from desired frictional outcomes. Unlike traditional trial-and-error and numerical optimization methods, our approach directly yields surface designs meeting specified frictional criteria with high accuracy and efficiency. This advancement in material surface engineering demonstrates the potential of machine learning in reducing the iterative nature of surface design processes. Our findings not only provide a new pathway for precise surface property tailoring but also suggest broader applications in material science where surface characteristics are critical.
\end{abstract}


\maketitle

\section{Introduction}
Traditional physics-based material design, often reliant on trial-and-error approaches, faces challenges of being resource-intensive and time-consuming \cite{casati2016}. To address these limitations, various iterative inverse design algorithms have emerged, introducing specific strategies to explore the configuration space efficiently.
Notably, Alex Zunger's group developed a genetic algorithm-based method for inverse design of materials' band structures, enabling the prediction of materials with specific band gaps \cite{dudiy2006,piquini2008,yu2012}. Additionally, a two-step inverse algorithm has been applied for tailoring the frictional properties of metals \cite{szeliga2006}. Recent advancements include the use of neural networks for rapid prediction of material properties, significantly accelerating the search process \cite{hanakata2018,yang2019,yu2019,hanakata2020,najafi2023}. Despite these advancements, challenges remain, particularly regarding the convergence and iteration requirements of these iterative optimization-based methods.

In this Letter, we demonstrate the use of a diffusion denoising probabilistic model (DDPM) for generating surfaces with specific frictional properties, leveraging advancements in machine learning \cite{ho2020,ho2022}. Our methodology includes training the DDPM on synthetic surfaces, designed using simplex noise and labeled with frictional properties determined from molecular dynamics simulations. Once trained, the model efficiently generates targeted surfaces with a specified frictional strength without further optimization. This approach aligns with recent trends in material science, where generative machine learning models have been employed for material design \cite{mao2020,chen2020} and stress prediction \cite{jiang2021,buehler2023}. It also parallels developments in physics, such as latent variational diffusion models for inverse problems in high-energy physics \cite{shmakov2023}. Our method is distinguished by its direct input of conditions into the model, generating outputs specific to those conditions.

\begin{figure}
    \centering
    \includegraphics[width=8.5cm]{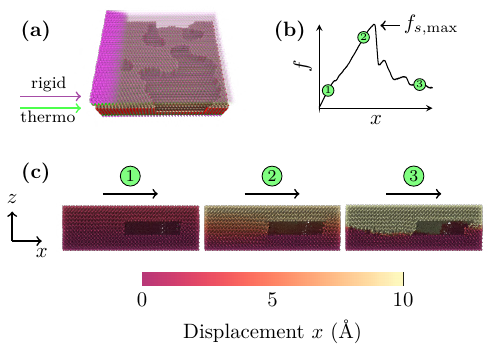}
    \caption{Molecular dynamics simulation. (a) We keep the uppermost and lowermost layers of the system rigid, and apply a Langevin thermostat to the atoms close to the layers. Here, the upper system surface is partly transparent to show the structure. (b) The lateral force is measured while the system is sheared, and the static friction is taken as the largest lateral force. (c) Cross-sectional view of the system initially (1), just before failure (2) and after failure (3).}
    \label{fig:fig1}
\end{figure}

\begin{figure*}
    \centering
    \includegraphics[width=\textwidth]{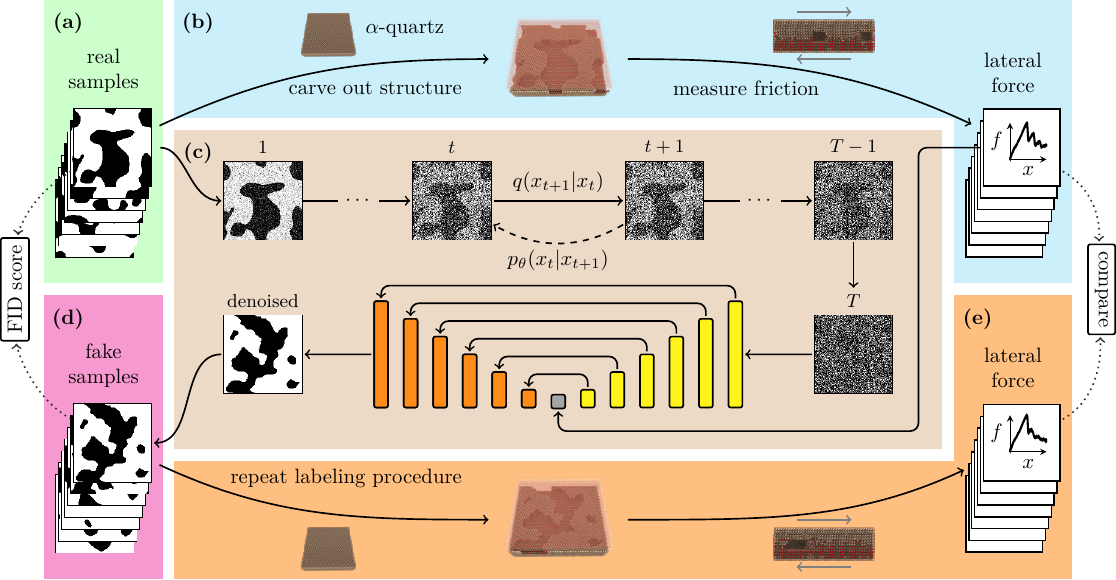}
    \caption{Training and validation of DDPM. (a) Binary structures are generated using simplex noise. (b) To label the structures by friction we carve out the structure in an $\alpha$-quartz crystal. The static friction is measured by shearing the system in molecular dynamics simulations. (c) Diffusion and denoising process. A DDPM is trained to gradually denoise an image. (d) Fake samples are generated by denoising noise using the DDPM. The quality of the generated images is evaluated by comparing to the real samples using the FID score. (e) Conditioning is validated by labeling fake samples generated for a given label and comparing the two labels. Illustrations of carved surfaces are rendered with Ovito \cite{stukowski2010}.}
    \label{fig:fig2}
\end{figure*}

\section{Methodology}
To prepare for training the DDPM, we first create a dataset linking surface topologies to friction properties. Simplex noise, known for its capability of generating realistic surfaces, is used to create these topologies \cite{perlin2002}. For each sample, we randomly select a scaling factor between 2 and 10 and choose a number of octaves from 1, 2, or 3. Lacunarity and persistence are fixed at 2 and 1, respectively. The noise is thresholded to ensure a uniform porosity of 40\% across all samples, represented as binary images ($128\times128$ pixels). For friction labeling, we perform MD simulations as displayed in Fig.~\ref{fig:fig1}. 
To do this, we carve out the structure of depth 2 nm in orthorhombic $\alpha$-quartz blocks with dimensions ($20\times20\times4$) nm$^3$. 
Thereafter, a perfect $\alpha$-quartz crystal of dimensions ($20\times20\times2$) nm$^3$ is put on top of the crystal, slightly shifted in $x$- and $y$-directions to avoid perfect sintering. 
During the simulation we apply a sandwich structure like seen in for instance Refs. \cite{li2014,nordhagen2023}, where the upper and lower 1 nm of the system is kept rigid and the atoms that are less than 1 nm from the rigid layers are controlled by a Langevin thermostat \cite{schneider1978}. A small normal pressure of 40 kPa is imposed on the rigid layers, while the entire system is sheared with a constant velocity of 5 m/s until the yield stress is reached, at a maintained temperature of 300K. Static friction is measured as the maximum lateral force on the upper surface, and samples are categorized into ten friction classes. Periodicity in the surface topologies is ensured for compatibility with the simulations. MD simulations are performed in LAMMPS \cite{thompson2022}, with the SiO$_2$ force field and parameters proposed by Broughton et al. \cite{broughton1997}.

In our work, we leverage the capabilities of generative neural networks, focusing on the diffusion denoising probabilistic model (DDPM), a variant of diffusion models first proposed by Ref. \cite{sohl-dickstein2015}. DDPM represents an advancement over initial diffusion models, showcasing ease of training and the ability to generate state-of-the-art images \cite{ho2020}.
The cornerstone of DDPMs lies in their ability to learn the reverse of a diffusion process -- a denoising operation -- by being exposed to images that incrementally increase in noise through a controlled diffusion process. The forward diffusion process is mathematically expressed as:
\begin{equation}
q(\boldsymbol{x}_{t+1}|\boldsymbol{x}_{t},\boldsymbol{x}_0)=\mathcal{N}(\boldsymbol{x}_{t};\boldsymbol{\mu}_t(\boldsymbol{x}_t,\boldsymbol{x}_0),\tilde{\beta}_t\boldsymbol{\mathbb{I}}),
\end{equation}
where $\boldsymbol{x}_t$ is the structure at time step $t$, $\boldsymbol{\mu}_t(\boldsymbol{x}_t,\boldsymbol{x}_0)$ and $\tilde{\beta}_t$, the mean and variance of the distribution, respectively, are derived from Brownian dynamics \cite{ho2020}:
\begin{align}
\boldsymbol{\mu}_t(\boldsymbol{x}_t,\boldsymbol{x}_0) &:= \frac{\sqrt{\bar{\alpha}_{t-1}}\beta_t}{1-\bar{\alpha}_t}\boldsymbol{x}_0 + \frac{\sqrt{\alpha_t}(1-\bar{\alpha}_{t-1})}{1-\bar{\alpha}_t}\boldsymbol{x}_t,\\
\tilde{\beta}_t &:= \frac{1-\bar{\alpha}_{t-1}}{1-\bar{\alpha}_t}\beta_t.
\end{align}
Here, the variances of the forward process, $\beta_t$, can be seen as hyperparameters, $\alpha_t:=1-\beta_t$ and $\bar{\alpha}_t:=\prod_{s=1}^t\alpha_s$.
The reverse process, in contrast, does not possess a closed-form solution and is inherently ambiguous. Neural networks is a good candidate to model such ambiguities, which is the main idea behind DDPMs. The denoising process is formulated as:
\begin{equation}
p_{\theta}(\boldsymbol{x}_{t-1}|\boldsymbol{x}_{t})=\mathcal{N}(\boldsymbol{x}_{t-1};\boldsymbol{\mu}_{\theta}(\boldsymbol{x}_{t}, t),\boldsymbol{\Sigma}_{\theta}(\boldsymbol{x}_{t}, t)),
\end{equation}
where the mean $\boldsymbol{\mu}_{\theta}(\boldsymbol{x}_t,t)$ and the variance $\boldsymbol{\Sigma}_{\theta}(\boldsymbol{x}_t,t)$, parameters of the model distribution, are determined by the neural network. Essentially, the network learns to estimate the conditional distribution of a less noisy image $\boldsymbol{x}_t$, given a noisier image $\boldsymbol{x}_{t+1}$, at each step of the reverse process. We refer to Ho et al. for more details \cite{ho2020}.

The conditional denoising model of choice was a U-Net backbone \cite{ronneberger2015} as suggested by Ref.~\cite{ho2020}, with 6 down-sampling and up-sampling blocks and skip-connections in between. An input block is used to increase the number of channels, and thus the complexity, while an output block is used to decrease the number of channels back to 1. Residual blocks \cite{he2015} are used to avoid vanishing gradients, while we use circular padding in our convolutional layers to favor periodic structures. Conditions are input into the upsampling blocks through embedding networks. We adapt the classifier-free conditioning idea from Ref.~\cite{ho2022}, blending outputs from models trained with and without conditions. Our network comprises a substantial 705 million trainable parameters. We refer to Appendix \ref{app:arch} for more details about the network architecture and Appendix \ref{app:cond} for information about conditioning and embedding.

During the training process, we utilized a learning rate of 0.0001, a batch size of 16, and employed 400 denoising steps. It was observed that reducing the number of denoising steps led to a marked decrease in image quality. The model's loss function is based on the mean-squared error, comparing the partial noise with the generated samples. To ensure equitable treatment of potential minority classes in the dataset, we implemented a weighted loss approach. The loss for each class is inversely proportional to its prevalence, calculated as $1/N_c$, where $N_c$ represents the count of samples in each class.

Post-training, the network generates artificial samples corresponding to specified static friction strengths. These samples are then re-labeled using molecular dynamics simulations, as previously detailed, and their accuracy is validated by comparing the generated friction strengths with the targeted values. This process of training and validation, illustrating the transition from theoretical modeling to practical application, is depicted in Fig.~\ref{fig:fig2}.

\begin{figure}
    \centering
    \includegraphics[width=8.5cm]{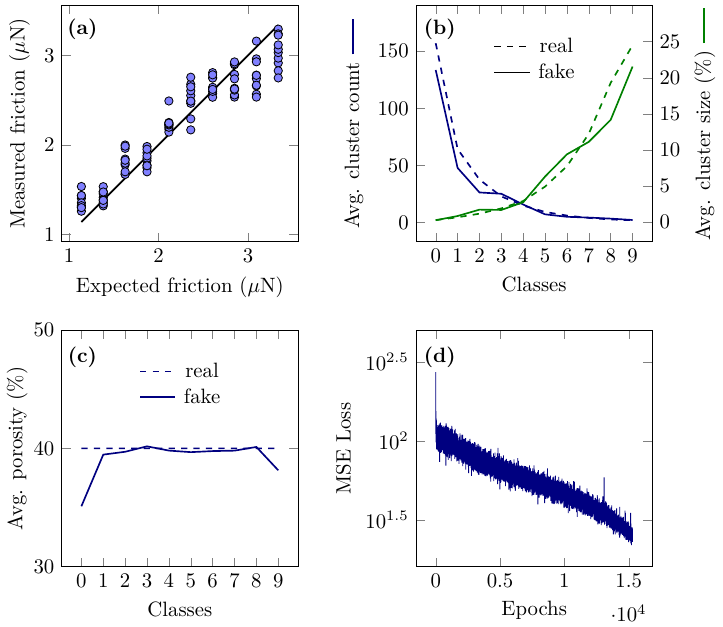}
    \caption{Performance of model. (a) The expected static friction of fake samples as a function of the measured friction. (b) Average number of clusters (blue lines) and average cluster sizes, given in peecentage of the total surface area (green lines) across classes. (c) Average porosity across all classes. (d) Mean-squared error loss as a function of epochs, log y-axis.}
    \label{fig:fig3}
\end{figure}

\section{Results}
Figure~\ref{fig:fig3} presents validations of our yet best model, which was trained on 4,000 samples for 15,000 epochs. Training the model took around 10 days on an NVIDIA A100 graphics card. The measured static friction of the fake samples agree with the expected static friction across all classes (Fig.~\ref{fig:fig3}a), with a mean-squared classification error of 0.50 $\mu$N$^2$. 45\% of the fake samples are classified correctly and 74\% fall into the correct class or neighbour classes. The morphology of the structures are analysed through a cluster and porosity analysis. In Fig.~\ref{fig:fig3}b, we compare average cluster size and number of clusters of the fake samples with the training dataset. An apparent observation is that the average cluster size is monotonically increasing with class, which seems to be the primary friction dependency when the contact area is fixed. The fake samples consistently follow the same trend and reproduce the training dataset almost perfectly both when it comes to the average cluster size and the average number of clusters in each class. The average porosity of the fake samples is compared to the ground truth (40\%) across the classes in Fig.~\ref{fig:fig3}c. Most of the fake samples have a porosity close to 40\%, but small deviations exist, especially for class 0 and 9. 
The mean-squared error loss is plotted as a function of epochs, which is strongly improving (Fig.~\ref{fig:fig3}d). Such a loss drop indicates that the model is still learning and improving.

\begin{figure}
    \centering
    \includegraphics[width=8.5cm]{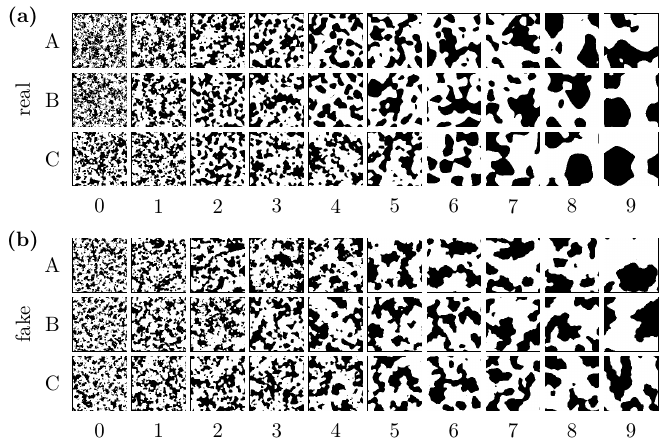}
    \caption{Example samples. (a) Real samples generated with simplex noise across the 10 classes. (b) Fake samples generated by the DDPM across the 10 classes.}
    \label{fig:fig4}
\end{figure}

Lastly, we display some real and fake example samples for all the classes in Fig.~\ref{fig:fig4}. At the first glance, the fake samples are very similar to the real one, but might have slightly sharper structures. For class 9, the average cluster size is slightly smaller for the fake samples compared to the real samples, which was also seen in Fig.~\ref{fig:fig3}b. This can be a minority class feature, even though the minority classes were weighted during training. Increasing the size of the training dataset will most probably improve sample generation of the minority classes.

\section{Conclusions and Perspectives}
In conclusion, we have  demonstrated a novel approach to the inverse design of surface properties, particularly focusing on friction, using a diffusion denoising probabilistic model (DDPM). This method marks a significant departure from traditional trial-and-error and numerical optimization techniques in surface engineering. By leveraging advanced machine learning algorithms, we have shown that it is possible to generate surface designs with precise frictional properties directly, thereby streamlining the design process and reducing the reliance on iterative testing.

Our methodology involved training a DDPM with a dataset of synthetic surfaces, each labeled with frictional properties derived from comprehensive molecular dynamics simulations. The resulting model was capable of producing high-fidelity surface designs that meet specific frictional criteria without the need for further optimization. This represents a significant efficiency improvement over conventional methods.
We obtain a mean-squared classification error of 0.50 $\mu$N$^2$ where 45\% of the generated surfaces fall into the correct class.

Moreover, the application of this approach extends beyond the scope of frictional surface design. The principles and techniques demonstrated here can be adapted for a wide range of material science applications, potentially revolutionizing how surface properties are tailored for various industrial needs.

As we move forward, there are opportunities to refine and expand this methodology. The potential to incorporate continuous conditions into the model \cite{ding2021}, adapt the network architecture for more complex problems, and explore other classes of generative models opens new avenues for research and application. These advancements could lead to even more precise control over material properties and further accelerate the pace of innovation in material science and engineering.

By demonstrating that the method works on a simple problem like this, we believe it will also work on more complex problems. For instance tailoring the color reflected from a surface \cite{andkjaer2014}. One can also move to non-binary samples and higher resolution. More complex problems might however require more complex network architectures. Our network is relative minimalist, and rather similar to the initial DDPM implementation \cite{ho2020}. The network can most likely be improved significantly by recent advances in machine learning. Other classes of generative models might also be examined, where a transformer network recently has shown impressive power \cite{peebles2023}. In the same work, it was shown that more complex architectures consistently provided better performance. The network used in this work is relatively shallow compared to the networks presented there because of memory limitations. Increasing the complexity, e.g. the number of features, would most likely improve the model.

\bibliographystyle{unsrtnat}
\bibliography{main}

\appendix

\begin{figure*}
    \centering
    \includegraphics[width=16cm]{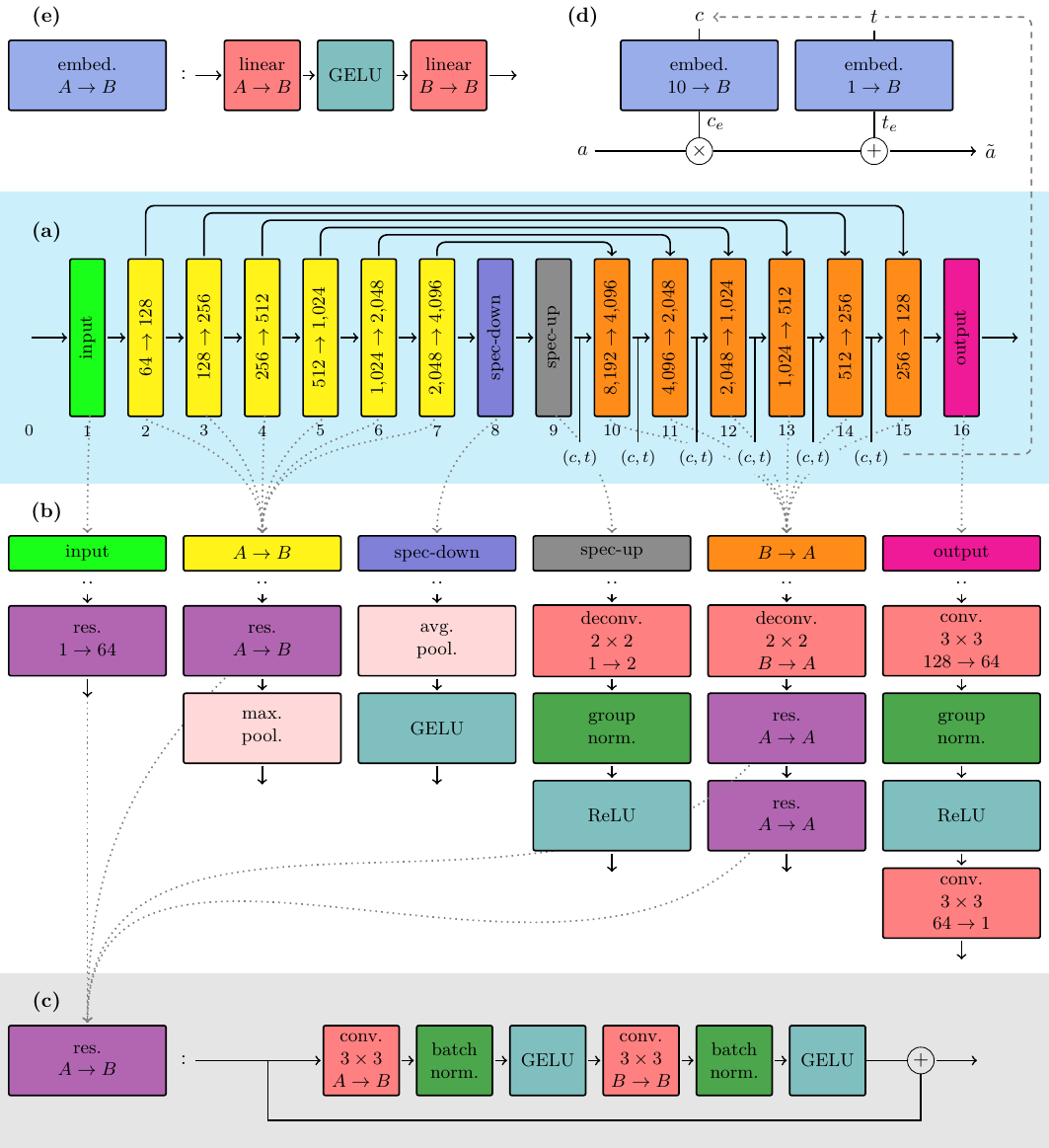}
    \caption{Architecture and conditioning of denoising model. (a) The denoising model takes a U-Net architecture, consisting of an input block (input), six standard downsampling blocks, one special downsampling block (spec-down), one special upsampling block (spec-up), six standard upsampling blocks and an output block (output). The information flow is shown by solid arrows: Information is passed linearly between the blocks, but skip-connections transport additional information directly to the standard upsampling blocks. (b) Each of the blocks consist of other blocks and layers. Both upsampling and downsampling blocks rely heavily on residual blocks (res), where the downsampling blocks also depend on deconvolutions. (c) Each residual block consists two stacked blocks of a convolutional layer and $3\times3$ kernel, batch normalization and GELU activation. (d) Embedding networks are used to input conditions and time step into upsampling blocks. (e) Embedding networks consist of a linear layer, GELU activation and then another linear layer.}
    \label{fig:fig5}
\end{figure*}

\section{Model Architecture} \label{app:arch}
The generative model that we use is based on a U-Net architecture \cite{ronneberger2015}, where the input resolution is gradually decreased down to a latent space while increasing the number of channels (downsampling). Thereafter, the latent space is upsampled back to the input shape, while taking inputs from the downsampling. The network consists of an input block, 6 general downsampling blocks, a special downsampling block converts input down to latent space, a special upsampling block from latent space, 6 general upsampling blocks and then an output block. There are skip-connections between pairwise upsampling- and downsampling blocks. The architecture blocks are illustrated in Fig.~\ref{fig:fig5}a and dimensionality details are given in Tab.~\ref{tab:tab5}.

Each of the 6 respective network building blocks consist of different subblocks and layers. The input block (input) consists of a residual block increasing the number of channels from 1 to 64. The downsampling blocks consist of a residual block doubling the number of channels, followed by a $2\times 2$ max pooling layer reducing the resolution by a factor of 2 in both directions. The special downsampling block (spec-down) applies average pooling, followed by Gaussian error linear unit (GELU) activation. The special upsampling block (spec-up) uses a deconvolutional layer with a $2\times 2$ kernel size, group normalization and rectified linear unit (ReLU) activation. Each up-sampling block consists of a $2\times2$ deconvolutional layer, followed by two residual convolutional blocks. Finally, the output block (output) consists of a $3\times 3$ convolutional layer, group normalization, ReLU activation and another $3\times 3$ convolutional layer. Each of the blocks are outlines in Fig.~\ref{fig:fig5}b.

Our residual block (res.) is structured following the principles outlined in Ref.~\cite{he2015}. It comprises a $3\times 3$ convolutional layer, batch normalization, and GELU activation, which is then repeated once more. A distinctive feature of this block is its use of a skip-connection that adds the input directly to the output, enhancing information flow through the network. The detailed architecture of this residual block is depicted in Fig.~\ref{fig:fig5}c.

\section{Conditioning and Embedding} \label{app:cond}
In our approach, conditions and timesteps are fed into all upsampling blocks, allowing the model to adjust its behavior dynamically during each denoising step. This is achieved by first converting the condition and timestep into a compatible dimensionality through embedding networks, which consist of a linear layer (for dimension transformation), GELU activation, and another linear layer (to increase variational parameters without changing dimensions). This process is shown in Fig.~\ref{fig:fig5}d. The transformed embeddings are then integrated with the output of the upsampling blocks using the linear equation 
\begin{equation}
    \tilde{\boldsymbol{a}}= \boldsymbol{a} \circ \boldsymbol{c}_e + \boldsymbol{t}_e,
\end{equation}
where $\boldsymbol{a}$ is the output of the previous block, and $\boldsymbol{c}_e$ and $\boldsymbol{t}_e$ are the embedded condition and timestep, respectively (Fig.~\ref{fig:fig5}e).

Additionally, we apply the classifier-free conditioning method from Ref.~\cite{ho2022}, which blends outputs from models trained with and without conditions. This is described by the equation 
\begin{equation}
    \tilde{\boldsymbol{\epsilon}}_{\theta}(\boldsymbol{z},c)=(1+w)\boldsymbol{\epsilon}_{\theta}(\boldsymbol{z},c)-w\boldsymbol{\epsilon}_{\theta}(\boldsymbol{z}),
\end{equation}
where $\boldsymbol{\epsilon}_{\theta}(\boldsymbol{z},c)$ and $\boldsymbol{\epsilon}_{\theta}(\boldsymbol{z})$ are outputs from conditional and unconditional models, respectively, and $w$ is a weight parameter balancing training fidelity and diversity. In our work, $w$ is set to 1 for all generated samples. In our diffusion model, the tensor $\boldsymbol{z}$ serves as the input, while $c$, representing the condition, is a scalar in this context.

\begin{table}
\centering
\caption{Output dimensionalities of the various blocks found in Fig.~\ref{fig:fig5}a. Upsampling blocks have twice as many channels as the corresponding downsampling blocks because of skip-connections.}
\label{tab:tab5}
\begin{tabular}{lllll}
\hline
block number & block ID & channels & x-dim & y-dim \\ \hline
0 & -- & 1 & 128 & 128 \\
1 & input & 64 & 128 & 128 \\
2 & down1 & 128 & 64 & 64 \\
3 & down2 & 256 & 32 & 32 \\
4 & down3 & 512 & 16 & 16 \\
5 & down4 & 1024 & 8 & 8 \\
6 & down5 & 2048 & 4 & 4 \\
7 & down6 & 4096 & 2 & 2 \\
8 & spec-down & 4096 & 1 & 1 \\
9 & spec-up & 8192 & 2 & 2 \\
10 & up1 & 4096 & 4 & 4 \\
11 & up2 & 2048 & 8 & 8 \\
12 & up3 & 1024 & 16 & 16 \\
13 & up4 & 512 & 32 & 32 \\
14 & up5 & 256 & 64 & 64 \\
15 & up6 & 128 & 128 & 128 \\
16 & output & 1 & 128 & 128 \\
\hline
\end{tabular}
\end{table}

\end{document}